\documentclass[aps, prx, reprint, groupedaddress,superscriptaddress, amsmath,amssymb,floatfix, nofootinbib,longbibliography]{revtex4-2}
\usepackage{graphicx}
\usepackage{dcolumn}
\usepackage{bm}
\usepackage{hyperref}
\usepackage{braket}
\usepackage{color}
\usepackage{comment}
\usepackage{sidecap}
\usepackage{bm}
\usepackage{xcolor}
\definecolor{BLUE}{named}{blue}

\begin{document}

\title{Small but large: Single organic molecules as hybrid platforms for quantum technologies}

\author{Burak Gurlek}
\email{burak.gurlek@mpsd.mpg.de}
\affiliation{Max Planck Institute for the Structure and Dynamics of Matter, Luruper Chaussee 149, 22761 Hamburg, Germany}

\author{Daqing Wang}
\email{daqing.wang@uni-bonn.de}
\affiliation{Institute of Applied Physics, University of Bonn,
Wegelerstr. 8, 53115 Bonn, Germany}

\begin{abstract}
Single organic molecules embedded in solid-state matrices exhibit remarkable optical properties, making them competitive candidates for single-photon sources and quantum nonlinear optical elements. However, the lack of long-lived internal states imposes significant constraints on their application in the broader field of quantum technologies. In this article, we reexamine the single-molecule host-guest system from first principles, elaborate on the rich internal states this system encompasses and put forward strategies to harness them for applications in quantum memory, spin-photon interface, spin register, and optomechanics. Further, we discuss the potential of leveraging the vast chemical space of molecules and highlight the challenges and opportunities for molecular systems along these directions.
\end{abstract}

\maketitle

\section{Introduction}
\label{sec:intro}

Molecular systems have emerged as promising platforms for quantum technologies due to their unique combination of structural tunability, rich internal degrees of freedom, and compatibility with solid-state integration. A notable example of their potential is the detection of a single pentacene molecule in a para-terphenyl crystal by Moerner et al. \cite{moerner-1989} and Orrit et al. \cite{orrit-1990}, which marked the beginning of significant advancements in quantum optics driven by polycyclic aromatic hydrocarbon (PAH) molecules~\cite{toninelli-2021, adhikari-2022}. Over time, the field of solid-state molecular quantum technologies has expanded to include a range of spin-based platforms, such as organometallic systems \cite{Bayliss-2020}, rare-earth molecular complexes \cite{Serrano2022} and organic radical species \cite{Gorgon2023}, driven by the versatility offered through chemical synthesis.

Among these systems, single PAH molecules are distinguished by their preeminent properties, including single-photon emission, close to Fourier-limited linewidth, high quantum efficiency, and excellent photostability, which make them excellent sources of indistinguishable photons \cite{lettow-2010, rezai-2018, duquennoy-2022}. These characteristics also enable single molecules to function as optical nonlinear media operating in the quantum regime, owing to the anharmonicity of their electronic excited states. This has been utilized to demonstrate a single-molecule optical transistor~\cite{hwang-2009} and frequency mixing between two optical beams~\cite{maser-2016,pscherer-2021}. Furthermore, the frequency of molecular photons can be tuned by the Stark effect~\cite{wild-1992, schaedler-2019,Duquennoy-2024}, exciting with a strong pump laser~\cite{colautti-2020a} or external strain~\cite{croci-1993, tian-2014,Fasoulakis-2023}. The fine control over the emission frequency facilitated the interfacing of remote molecules through single photons\,\cite{rezus-2012,wang-2019}. Moreover, PAH molecules are compatible with photonic structures such as dielectric~\cite{lee-2011,Checcucci-2016} and plasmonic antennas~\cite{zirkelbach-2020}, optical waveguides~\cite{faez-2014,skoff-2018,grandi-2019,colautti-2020b,ren-2022} and on-chip resonators~\cite{rattenbacher-2023}. Recent experiments have demonstrated the strong coupling of a single molecule to a Fabry-P\'erot microcavity~\cite{wang-2019, pscherer-2021}. Having frequency tunability and controllable doping concentration, near-field couplings between two closely spaced molecules via coherent dipole-dipole interactions have also been shown~\cite{hettich-2002,trebbia-2021,lange-2024}. In addition, the narrow-linewidth optical transition makes single molecules ultra-sensitive probes for their local environment, e.g., charge fluctuations\,\cite{shkarin-2021}, and mechanical oscillations \cite{puller-2013, tian-2014}.  

Spectroscopic detection of single PAH molecules relies on their insertion as impurities into a host material, i.e., forming a \emph{host-guest system} through van der Waals interactions. With suitable choices of host-guest combinations, such a system can feature two favorable properties. First, the van der Waals interaction allows the electronic excitations of the guest to remain largely localized within its own molecular orbitals, with only weak perturbation from the host. In this regard, the guest's ground state and lowest electronic excited state facilitate an optical transition, which could feature a nearly lifetime-limited linewidth at liquid helium temperatures, on par with atomic transitions in vacuum. Second, this transition can be approximated as a two-level cycling transition, where population decays to other levels are recycled through interactions with the matrix, e.g., fast vibrational relaxation, allowing measurements using only a single laser beam. It is worth noting that the success of most past achievements in molecular quantum optics builds primarily upon the efficient isolation of such a two-level transition. Beyond this simplified two-level picture, the molecular host-guest system represents a hybrid quantum system, incorporating electronic, nuclear, and spin degrees of freedom. The presence of this multitude of couplings has been considered detrimental, as they introduce incoherent leakage channels.

In this article, we introduce single molecules embedded in a solid-state host as a hybrid quantum system, provide a detailed discussion of their various degrees of freedom, and outline the prospects of harnessing them for applications that have not been realized on molecules so far. The article is organized as follows. In section\,\ref{sec:host-guest}, we examine the molecular host-guest system from first principles, revisit its wavefunction, energy landscape, and associated transitions. A particular focus will be given to the quantum nature of its vibrational and spin degrees of freedom. In section\,\ref{sec:qt}, we discuss the potential of using vibrational and spin states for molecular quantum memory, optomechanics, spin-photon interfaces and chemically configurable spin registers. We will further expand the discussion towards exploring the vast chemical space of molecules for quantum applications. Finally, we discuss the theoretical and experimental challenges associated with these approaches in section\,\ref{sec:challenges} and conclude with section\,\ref{sec:discussion}.

\begin{figure}[!h]
\begin{center}
\includegraphics[width=0.98\columnwidth]{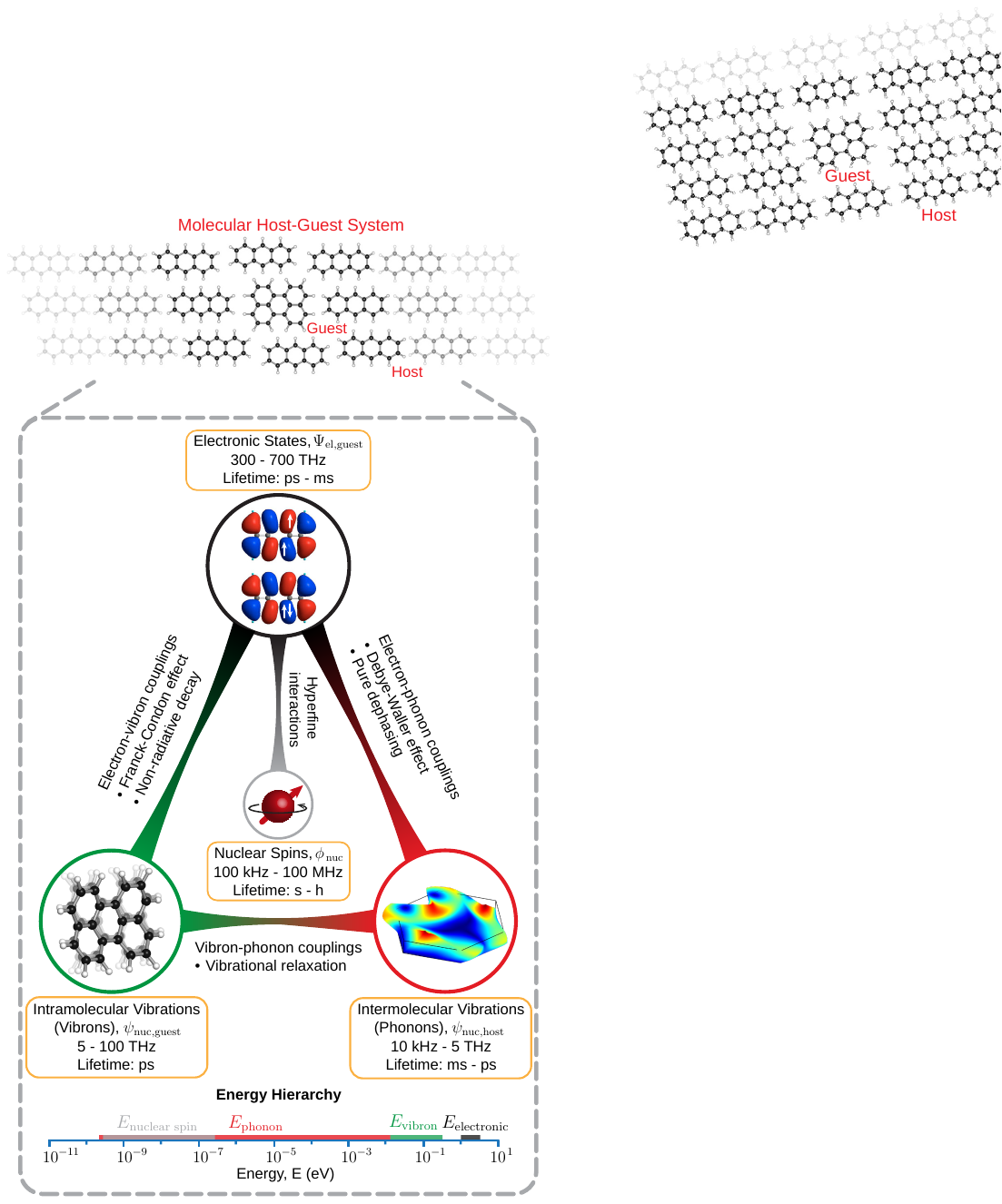}
\caption{Upper panel: Illustration of a molecular host-guest system, specifically shown here for anthracene crystal hosting a perylene molecule. Lower Panel: Schematics for the relevant states and their couplings. The guest's spin-orbital states (top) interact with its own intramolecular vibrations (lower left) and intermolecular vibrations of the host crystal (lower right) via electron-vibron and electron-phonon couplings, respectively. Electronic states are further coupled to nuclear spin degrees of freedom via hyperfine interactions (middle). The vibrational states are coupled to each other via vibron-phonon couplings. The energy hierarchy of these states is shown at the bottom of the lower panel, where $E_\text{electronic}\sim$~1~eV~-~3~eV, $E_\text{vibron}\sim$~20~meV~-~0.5~eV, $E_\text{phonon}\sim$~0.3 neV~-~20~meV, $E_\text{nuclear spin}\sim$~$0.4$~neV~-~$400$~neV. The quoted energy scale for nuclear spins refers to their range of hyperfine coupling strength.}
\label{fig:couplings}
\end{center}
\end{figure}

\section{Emerging Insights into Molecular Host-Guest Systems}
\label{sec:host-guest}
Molecules have emerged as engineerable building blocks for future quantum technologies, thanks to their rich internal structures and the vast diversity offered by the chemical space~\cite{tesch-2002,Albert-2020,toninelli-2021,Augenbraun2023}. Figure\,\ref{fig:couplings} illustrates the multiple degrees of freedom in the molecular host-guest system and their couplings. While there is also growing interest in exploiting these internal states~\cite{gurlek-2021,Mena-2024,steiner2024}, earlier studies have predominantly focused on the electronic degree of freedom~\cite{toninelli-2021}, leaving other intriguing aspects largely unexplored. To highlight the promise of molecular host-guest systems for future explorations in quantum science, we revisit these ``overlooked'' internal states. Drawing inspiration from previous works~\cite{Diestler-1979}, we perform a first-principle analysis of molecular host-guest systems to explore their quantum mechanical characters and interactions in detail.

\subsection{Revisiting the Host-Guest Wavefunction}
{The wavefunction of a molecular host-guest system can be expressed following the Born-Oppenheimer approximation similar to the case for a single molecule as
\begin{align}
    \Psi\approx&\Psi_\text{el}(\mathbf{r},\mathbf{s};\mathbf{R},\mathbf{I})\Psi_\text{nuc}(\mathbf{R},\mathbf{I}),\label{eq:mol_wavefuc}
\end{align}
where the terms on the right-hand side represent the many-body electronic wavefunction of $n_\text{e}$ electrons of all host and guest molecules with spatial $\mathbf{r}=(\mathbf{r}_1,\mathbf{r}_2,\ldots,\mathbf{r}_{n_\text{e}})$ and spin coordinates $\mathbf{s}=(\mathbf{s}_1,\mathbf{s}_2,\ldots,\mathbf{s}_{n_\text{e}})$, and the many-body nuclear wavefunction of $n_\text{n}$ nuclei of all host and guest molecules with spatial $\mathbf{R}=(\mathbf{R}_1,\mathbf{R}_2,\ldots,\mathbf{R}_{n_\textbf{n}})$ and spin coordinates $\mathbf{I}=(\mathbf{I}_1,\mathbf{I}_2,\ldots,\mathbf{I}_{n_\text{n}})$, respectively. Typically, a large-band-gap host material is chosen such that its lowest electronic excitation is energetically higher than the states of interest in the guest. In this case, the electronic excitation of the guest remains well isolated from the host. It is thus possible to separate the electronic degrees of freedom such that

\begin{align}
\Psi_\text{el}\approx\Psi_\text{el,guest}(\mathbf{r}_\text{g},\mathbf{s}_\text{g};\mathbf{R},\mathbf{I})\Psi_\text{el,host}(\mathbf{r}_\text{h},\mathbf{s}_\text{h};\mathbf{R},\mathbf{I})\,,
\end{align}
where $\mathbf{r}_\text{g}$, $\mathbf{s}_\text{g}$ ($\mathbf{r}_\text{h}$, $\mathbf{s}_\text{h}$) denote the spatial and spin coordinates of the guest (host), respectively. When studying low-lying excitations in the guest, the electronic states of the host are unaffected and can be traced out~\footnote{The host's electronic degrees of freedom dress that of the guest molecule, and play an important role in determining the geometry of the guest molecule in the solid state.}. Hence, the system's electronic wavefunction reduces to the spin-orbital state of the guest $\Psi_\text{el,guest}$, as shown by the top node of the triangular diagram in Fig.\,\ref{fig:couplings}.

Considering the nuclear wavefunction, the interaction between the spin state and the mechanical motion of the nuclei is relatively weak compared to the dominant electrostatic interactions that govern nuclear motion, it is thus feasible to separate these two degrees of freedom following
\begin{align}
    \Psi_\text{nuc}(\mathbf{R},\mathbf{I})\approx \psi_\text{nuc}(\mathbf{R})\phi_\text{nuc}(\mathbf{I}).
\end{align}
Moreover, decoupling of the host and guest molecular orbitals allows the decomposition of the nuclear wavefunction into those of the host and the guest. The vibrational part of the nuclear wavefunction can be expressed in terms of the normal modes of the guest molecule and the host matrix with coordinates $\mathbf{Q}$ and $\mathbf{u}$ as~\cite{Diestler-1979,gurlek-2023-thesis}
\begin{align}
    \psi_\text{nuc}(\mathbf{R})\approx \psi_\text{nuc,guest}(\mathbf{Q})\psi_\text{nuc,host}(\mathbf{u})\,,
\end{align}
where $\psi_\text{nuc,guest}(\mathbf{Q})$ and $\psi_\text{nuc,host}(\mathbf{u})$ denote the intramolecular guest and intermolecular host vibrational wavefunctions,~respectively (see the bottom nodes of the triangular diagram in~Fig.\,\ref{fig:couplings}.) We note that the separation between host and guest vibrational degrees of freedom breaks down for low-frequency guest modes, which appear in large PAH molecules. These modes are discussed in detail in Section~\ref{sec:intramolecular_vib}. In the remainder of the text, we use the terms \emph{vibron} for the quantized intramolecular normal modes of guest molecule and \emph{phonon} for the quantized intermolecular normal modes of host matrix. 

As a result, the wavefunction of the host-guest system can be approximated as 
\begin{align}
\Psi\approx&\Psi_\text{el,guest}(\mathbf{r}_\text{g},\mathbf{s}_\text{g};\mathbf{R},\mathbf{I})\psi_\text{nuc,guest}(\mathbf{Q})\nonumber\\ &\times \psi_\text{nuc,host}(\mathbf{u})\phi_\text{nuc}(\mathbf{I}),
\end{align}
with their order reflecting the hierarchy of energy scales associated with these states, as indicated in the bottom panel of Fig.~\ref{fig:couplings}.

Excitations across these various degrees of freedom result in the complex energy level structure depicted in Fig.~\ref{fig:energy_levels}(a). Depending on the total electron spin quantum number $S$, the guest's electronic state $\Psi_\text{el,guest}$ can be represented as $\ket{S_0}, \ket{S_1},...$ for \textit{singlet states} with $S=0$, or $\ket{T_1}, \ket{T_2},...$ for \textit{triplet states} with $S=1$. The energy levels in Fig.~\ref{fig:energy_levels}(a) can be thus grouped into singlet and triplet manifolds based on $\Psi_\text{el,guest}$. We focus primarily on three, namely the ground state ($S_0$), lowest-singlet excited state ($S_1$) and lowest-triplet excited state ($T_1$) manifolds, as they play the most important role in molecular quantum optics.

Each manifold encompasses a range of vibron and phonon states, which can be written as $\ket{\bm{\nu}_v,\bm{\nu}_p}$ in the Fock state representation of the vibron and phonon wavefunctions. Here, $\ket{\bm{\nu}_v}=\Pi_i\ket{\nu_v^{(i)}}$ and $\ket{\bm{\nu}_p}=\Pi_i\ket{\nu_p^{(i)}}$ denotes the occupation of vibron and phonon modes with vibrational quantum numbers $\nu_v^{(i)}$ and $\nu_p^{(i)}$ for the $i^\text{th}$ vibron and phonon modes, respectively. In addition, the many-body nuclear spin states can be expressed as $\phi_\text{nuc}(\mathbf{I})=\sum_{\{m_{I_i}\}} C_{\{m_{I_i}\}}\bigotimes_{i=1}^{n_\text{n}}\ket{I_i,m_{I_i}},$
which denotes a linear superposition of basis states formed by the tensor products of uncoupled spin state $\ket{I_i,m_{I_i}}$ from $n_\text{n}$ nuclei, with the corresponding coefficients $C_{\{m_{I_i}\}}=C_{m_{I_1}, m_{I_2},...,m_{I_{n_\text{n}}}}$. Here, $I_i$ and $m_{I_i}$ denote the spin quantum numbers of the $i^\text{th}$ nucleus. As a result, the wavefunction stemming from Eq.~\eqref{eq:mol_wavefuc} can be represented as 
\begin{align}
    \Psi\approx\ket{S(T)_j;\bm{\nu}_v,\bm{\nu}_p;I_i,m_{I_i}}\,,\label{eq:mol_wavefuc_ket}
\end{align}
where $j$ is the integer index of the spin-orbital state and $S(T)$ denotes its spin multiplicity (singlet or triplet). In the following discussion of electronic and vibronic transitions, we will first omit the nuclear spin states, as they primarily manifest through hyperfine interactions with electron spins and are relevant only when probing the triplet manifold, such as in optically detected magnetic resonance experiments. We will revisit these states in Section\,\ref{sec:spin}.

\begin{figure}[ht]
\begin{center}
\includegraphics[width=0.99\columnwidth]{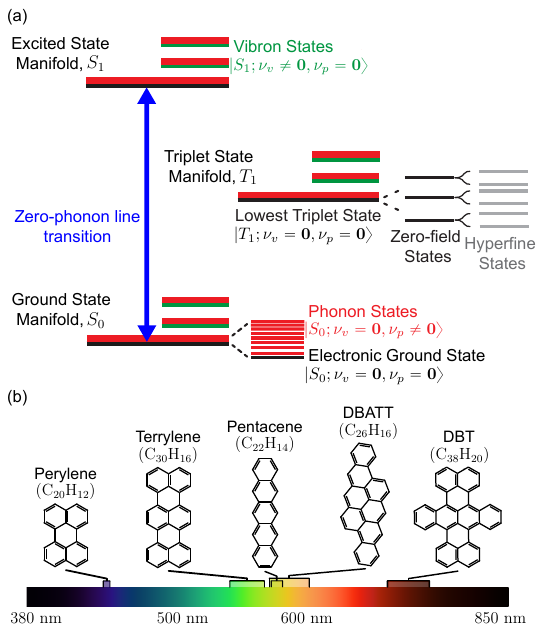}
\caption{(a) The energy levels of a single molecule embedded in a matrix. The state $\ket{S_{0(1)};\bm{\nu}_v,\bm{\nu}_p}$ belongs to the ground (excited) singlet state manifold and has vibrational occupations in the intramolecular $\ket{\bm{\nu}_v}=\Pi_i\ket{\nu_v^{(i)}}$ and intermolecular vibrational modes $\ket{\bm{\nu}_p}=\Pi_i\ket{\nu_p^{(i)}}$. The indices $\nu_{v(p)}^{(i)}=0,1\ldots$ denote the vibrational quantum number of the $i^{\text{th}}$ vibron (phonon) mode. Similarly, $\ket{T_{1};\bm{\nu}_v,\bm{\nu}_p}$ represents the states in the triplet state manifold. (b) PAH molecules commonly used in quantum optics experiments and their range of approximated zero-phonon-line wavelengths~\cite{adhikari-2023}.}
\label{fig:energy_levels}
\end{center}
\end{figure}

\subsubsection{Ground and excited states}
The ground state of the system $\ket{S_0;\bm{\nu}_v=\mathbf{0},\bm{\nu}_p=\mathbf{0}}$ is a singlet state, in which all $\pi$-orbitals are filled and all electron spins are paired. The excitation of one electron to a $\pi^*$ orbital leaves the molecule in the lowest singlet excited state $\ket{S_1;\bm{\nu}_v= \mathbf{0},\bm{\nu}_p=\mathbf{0}}$. Alternatively, if this excitation is accompanied by a spin-flip, the molecule reaches the lowest triplet excited state $\ket{T_1;\bm{\nu}_v= \mathbf{0},\bm{\nu}_p=\mathbf{0}}$. The energy of $\ket{T_1;\bm{\nu}_v= \mathbf{0},\bm{\nu}_p=\mathbf{0}}$ is typically lower than that of $\ket{S_1;\bm{\nu}_v= \mathbf{0},\bm{\nu}_p=\mathbf{0}}$ by the exchange energy. The transition energies of these two excited states to the ground state lie in the optical frequency domain and range from ultraviolet (UV) to near-infrared (NIR) wavelengths, depending on the size and geometry of the molecule. The triplet state $\ket{T_1;\bm{\nu}_v= \mathbf{0},\bm{\nu}_p=\mathbf{0}}$ consists of three sublevels, resulting from the anisotropic magnetic dipole-dipole interactions of two unpaired electrons, which remain non-degenerate even in the absence of an external magnetic field. This is called \textit{zero-field splitting}\,\cite{turro-2010-book}. For PAH molecules emitting in the UV to visible range, the zero-field splitting can reach a few gigahertz\,\cite{miyokawa-2024}. Moreover, coupling of the electron spin to nuclear spins in the molecule and the matrix leads to \textit{hyperfine splittings}, which lie in the radio-frequency range.

In addition, each manifold includes vibron states $\ket{S(T),\bm{\nu}_v\neq \mathbf{0},\bm{\nu}_p=\mathbf{0}}$ corresponding to librational and bond vibrations of the guest molecule, with frequencies that can reach up to $100$~THz. The coupling between electronic and intramolecular vibrational states (electron-vibron coupling) gives rise to the well-known Franck-Condon physics, where the displacement of the excited state potential energy surface along a normal-mode coordinate enables transitions between vibronic states in different manifolds. The transition probabilities, i.e., \textit{Franck-Condon factors}, are related to the Huang-Ryhs factor of each normal mode~\cite{reitz-2020}. For certain PAH molecules, the simple Franck-Condon picture can be further complicated by the Herzberg-Teller effect, where couplings between the electronic and nuclear degrees of freedom are not parametric~\cite{osad-2013,kong-2021}. Furthermore, non-adiabatic couplings between the electronic and nuclear degrees of freedom result in non-radiative decay from the excited state manifold via electron-vibron couplings~\cite{englman-1970,bassler-2024}. This could lead to non-unity quantum efficiencies of photon emission~\cite{erker-2022}.

Besides, phonon states $\ket{S(T),\bm{\nu}_v= \mathbf{0},\bm{\nu}_p\neq \mathbf{0}}$ typically fall within the $10$~kHz~-~$5$~THz range and lead to the deformation of molecular orbitals. The resulting electron-phonon couplings yield various optical transitions between different manifolds involving vibrons and phonons, similar to the Franck-Condon physics discussed earlier. Depending on the frequency range, these vibrational states exhibit lifetimes in the range of milliseconds to picoseconds due to different vibrational relaxation mechanisms~\cite{srivastava-1990}. 

\subsubsection{Zero-phonon-line transition}
Despite the complexity of the energy structure shown in Fig.\,\ref{fig:energy_levels}(a), decay to the ground state manifold typically occurs from the lowest singlet excited state $\ket{S_1;\bm{\nu}_v=\bm{0},\bm{\nu}_p=\bm{0}}$ due to fast vibrational relaxation. This is known as \textit{Kasha's rule}\,\cite{kasha-1950}. The most prominent of such transitions occurs between $\ket{S_1;\bm{\nu}_v=\bm{0},\bm{\nu}_p=\bm{0}}$ and the ground state $\ket{S_0;\bm{\nu}_v=\bm{0},\bm{\nu}_p=\bm{0}}$ as a result of large wavefunction overlap, i.e., small Huang-Rhys factors of vibron and phonon states. This transition is referred to as the \textit{$00$-zero-phonon line}, abbreviated as ZPL in the following discussion. Due to coupling with the vibrational degrees of freedom, the ZPL transition strength is reduced compared to that of a perfect two-level system by a factor related to the Franck-Condon factors of each intramolecular and intermolecular vibrational modes in the host-guest system. The latter is called the \textit{Debye-Waller factor} and is temperature-dependent due to low-frequency phonons involved~\cite{reitz-2020}. For a number of PAH molecules [see Fig.~\ref{fig:energy_levels}(b)], the ZPL transition lies in the visible to near-infrared frequency range, making them favorable for quantum optics experiments. At liquid helium temperatures and with appropriately chosen host matrices, the ZPL transition can narrow to near its lifetime limit. This benefits from the non-polar nature of the electronic wavefunction. It is worth noting that the insertion into a host matrix can slightly break the centrosymmetry, thereby inducing a net electric dipole\,\cite{smit-2024} and altering the vibrational mode structure~\cite{zirkelbach-2021}. Above liquid helium temperatures, the electron-phonon coupling leads to a reduction in the ZPL strength in addition to a decrease in the Debye-Waller factor as a result of elastic electron-phonon scatterings, known as \textit{pure dephasing}~\cite{skinner-1988}.

In summary, a single molecule embedded in a solid-state host exhibits a rich energy structure supporting a multitude of well-defined transitions ranging from radio to optical frequencies. Next, we delve into details of these molecular degrees of freedom and discuss their quantum mechanical characters.

\subsection{Vibrational States}\label{label:VibStates}
\label{sec:vibration}
The optomechanical character of molecules was first discovered by Raman~\cite{raman-1928}, as the optical excitation of molecules yields light scattering to new frequencies determined by the intramolecular vibrational energies. In the context of molecular quantum optics, this optomechanical character has traditionally been regarded as a source of decoherence due to fast vibrational relaxation processes. Yet, if vibrational states can be engineered to have long lifetimes, they could be utilized as a resource in quantum technologies.

\begin{figure*}[t]
\centering
\includegraphics[width=13cm]{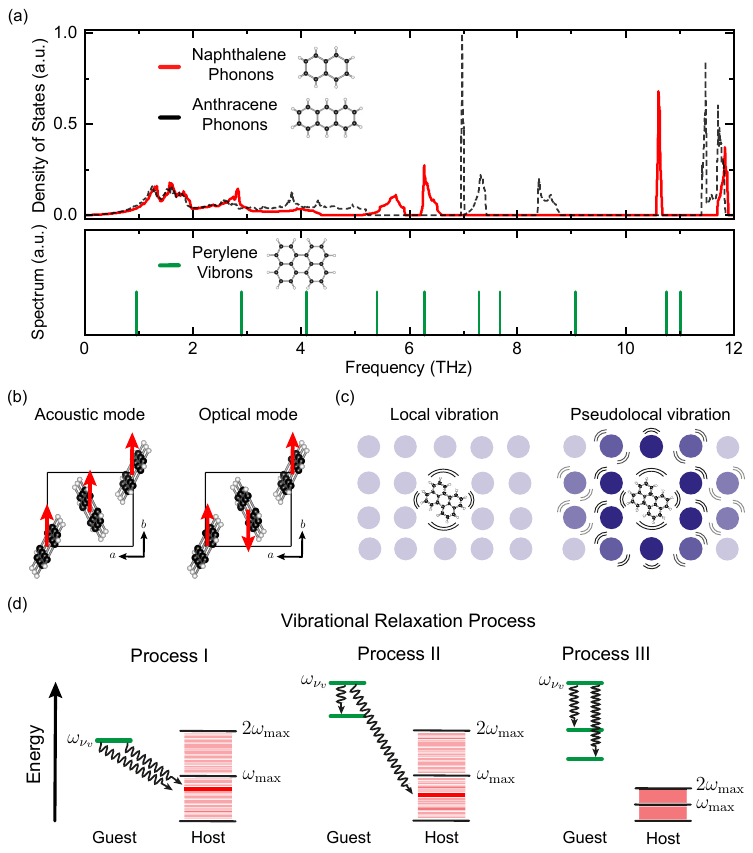 }
\caption{(a) Phonon density of states of naphthalene and anthracene crystals (top), and intramolecular vibrations of a perylene molecule (bottom), shown by red, black, and green lines, respectively, plotted up to $12$~THz. For clarity, only $10$ out of $90$ vibrational modes of perylene that lie in this spectral range are shown. The \emph{ab initio} calculations are done using Vienna Ab initio Simulation Package~\cite{kresse1993ab,kresse1994ab,kresse1996efficiency,kresse1996efficient} with Perdew-Burke-Ernzerhof functional~\cite{Perdew-1996} and Tkatchenko-Scheffler dispersion correction~\cite{tkatchenko2009accurate}. We used the Phonopy python package~\cite{phonopy-phono3py-JPSJ} to calculate the phonon density of states for a $2 \times 4 \times 2$ supercell. (b) Schematic of acoustic and optical phonons in an anthracene crystal. The crystal axes are denoted by $a$ and $b$. (c) Illustrations of local and pseudolocal modes in the host-guest system. (d) Vibrational relaxation mechanisms for an intramolecular vibration of a guest molecule in a host environment. The symbols $\omega_{\nu_v}$ and $\omega_\text{max}$ denote the intramolecular vibrational frequency of the state $\ket{S_0;[0,\ldots,0,1,0\dots,0],\bm{\nu}_p=\bm{0}}$  and maximal phonon frequency in the host matrix, respectively.}
\label{fig:Fig_Vibrations}
\end{figure*}

\subsubsection{Intermolecular Vibrational States}\label{sec:intermolecular_vib}
The typical vibrational landscape of a molecular host-guest system is shown with an example of a perylene-doped naphthalene and anthracene crystal in Fig.~\ref{fig:Fig_Vibrations}(a), where the phonon density of states in naphthalene crystal (red lines), anthracene crystal (black lines) and intramolecular vibrations of a perylene molecule (green vertical lines) are plotted up to $12$~THz. Among many mechanical degrees of freedom, phonons corresponding to the center-of-mass translational motion of host molecules are the lowest in energy scale. These modes approximately span the frequency range $10$~kHz~-~$4.5$~THz with a cut-off frequency determined by the unit-cell size of the host molecular crystal, i.e., around $1$~nm for PAH molecular crystals. Assuming molecules as rigid bodies, there are $3N-6$ lattice modes for $N$ number of molecules in the crystal, which practically renders the spectrum of such modes as a continuum. These modes can be classified as acoustic and optical phonons based on their properties at the zero-phonon wavevector. Acoustic phonons result from in-phase motion of molecules as rigid bodies~[see Fig.~\ref{fig:Fig_Vibrations}(b)] and their frequencies range in between $0-1.5$~THz for PAH molecular crystals. These modes possess linear dispersion for low frequencies, i.e., up to $\sim 100-200$~GHz, which results in a small density of states. The lifetime of these modes is determined by absorption and boundary scattering effects and can reach up to millisecond timescales for modes up to megahertz frequencies~\cite{srivastava-1990}. However, the electron-phonon couplings to these modes are small due to large phonon mode volumes in the molecular crystal~\cite{gurlek-2021}. Hence, optically accessing these modes is difficult.

Besides, high-frequency acoustic phonons and optical phonons have nonlinear dispersion relations with low group velocities, and correspond to out-of-phase motion of molecules in the unit-cell [see Fig.~\ref{fig:Fig_Vibrations}(b)]. These modes typically have lifetimes in the order of picoseconds due to the decay into acoustic modes at lower frequencies~\cite{klemens-1966}. They have considerable electron-phonon coupling strengths and hence constitute approximately $10\%$ of the overall light scattering for typical host-guest systems at liquid helium temperatures~\cite{reitz-2020}. The transitions to these states are visible in the phonon-wing spectra whose probability peaks at around $1.5-2$~THz and gradually decays towards the maximal phonon frequencies. Such a low optical excitation probability at high frequencies stems from low phonon DOS and the size of the guest molecule, i.e., its electron density in $k$-space~\cite{nazir-2016}. Due to their high densities and fast vibrational relaxations, these states are typically considered as a bath in molecular quantum optics~\cite{gurlek-2021}.

The phonon density of states possesses several other band structures at higher frequencies compared to $\omega_\text{max}$, i.e., $>4.5$~THz, which corresponds to the librational and rotational motions of individual lattice molecules~\cite{zhang-2014}. For anthracene and larger PAH molecular crystals, low-frequency phonons of such character are amalgamated into the continuum of phonons, and extend their cut-off frequency~\cite{dlott-1986}, as shown in Fig.~\ref{fig:Fig_Vibrations}(a). As argued above, the electron-phonon couplings to such modes are less probable due to the cut-off frequency induced by the electron density in $k$-space and hence play a minor role in molecular host-guest systems.

\subsubsection{Intramolecular Vibrational States}\label{sec:intramolecular_vib}
The molecular host-guest system additionally possesses intramolecular vibrations of the guest molecule [see the lower panel of Fig.~\ref{fig:Fig_Vibrations}(a)]. There are $3n-6$ vibrations corresponding to $n$ number atoms in the molecule, and they correspond to librational and bond vibrations of the guest molecule, reaching up to $93$~THz for the C-H stretching motion. Frequencies of librational modes are determined by the size of the molecule. The lowest libration frequency for perylene molecule in vacuum is $0.9$~THz. In the range of $0-5$~THz, librational modes of the guest molecule hybridize with the translational lattice modes. This results in slightly delocalized vibrational states over the host molecules~[see Fig.~\ref{fig:Fig_Vibrations}(c)], so-called \emph{pseudolocal phonon modes}~\cite{Fleischhauer-1992}. Depending on the insertion of the guest molecule, some of these modes are Raman (Franck-Condon) active, and due to the semi-localized nature of their vibration over the guest molecule, they can have large electron-vibron coupling strengths compared to phonon modes of the host matrix. This is evident in the single molecule emission spectra, where the phonon-wing spectrum shows pronounced Lorentzian peaks below $5$~THz with lifetimes in the order of picoseconds~\cite{kummer-1997}. In addition, these modes are shown to be responsible for temperature-dependent line broadening of the ZPL transition~\cite{skinner-1988}. 

The rest of the intramolecular vibrations are localized on the guest molecule [see Figs.~\ref{fig:Fig_Vibrations}(c)] as the host vibrational density of states is sparse. Hence, these modes have very large electron-vibron coupling strengths in the order of their frequencies~\cite{zirkelbach-2021,gurlek-2023-thesis}. Such ultra-strong couplings can be witnessed from the emission and absorption spectrum of the host-guest system.

\subsubsection{Vibrational Relaxation}
Despite long radiative lifetimes of molecular vibrations in the gas phase that range up to seconds~\cite{brueck-1976,campbell-2008}, the nonradiative vibrational relaxation in a solid typically yields lifetimes in the order of picosecond irrespective of the frequency and temperature range~\cite{zirkelbach-2021}. This process stems from the anharmonicity of the potential energy surface along host and guest vibrational coordinates, and results in short vibrational lifetimes even though certain vibrations of the guest molecule are in principle mechanically shielded due to gaps in the host vibrational band, as seen within the $7-11$~THz spectral region of the phonon density of states and perylene vibrational spectrum in Fig.~\ref{fig:Fig_Vibrations}(a). Based on the perturbative treatment of this problem, the vibrational relaxation process due to cubic anharmonic interactions between vibrational modes can be divided into three categories~\cite{hill-1988} depending on the frequency of the guest vibrational mode,~$\omega_{\nu_v}$, as shown in Fig.~\ref{fig:Fig_Vibrations}(d). In the first process, the guest molecule's vibrational frequency is smaller than two times the phonon cut-off frequency,~$\omega_\text{max}$, i.e., cut-off frequency of the two-phonon density of states, such that the vibration of the guest molecule relaxes via two lattice phonons based on energy and momentum conservation as a result of the three-body interaction between the normal-mode displacements. The second process happens when the guest molecule's vibrational frequency lies above the cut-off frequency of the two-phonon density of states, where an intramolecular vibration of the guest molecule could assist the non-radiative relaxation. If the guest molecule's vibrational frequency is too large compared to the phonon cut-off frequency, the vibrational relaxation occurs via intramolecular relaxation in the guest molecule as the third process. Besides, as this picture is perturbative, the higher-order anharmonicities between vibrational modes also play a role once the energy conservation cannot be satisfied via relaxation to two vibrations. As a result, despite the weak couplings between the intramolecular guest vibrations and phonon modes, the immense number of decay paths provided by the continuum of phonon modes yield picosecond-scale vibrational lifetimes. 
Even with such short lifetimes, these vibrations are mostly in their quantum ground state even at room temperature, an interesting property that could be exploited to push optomechanical quantum technologies beyond millikelvin  temperatures~\cite{tesch-2002}. 

\subsection{Spin states}
\label{sec:spin}
The spin states of electrons and nuclei in the molecular host-guest system offer additional parameter spaces that hold potential for quantum applications. We start by introducing the properties of the lowest triplet state $\ket{T_1;\bm{\nu}_v=\bm{0},\bm{\nu}_p=\bm{0}}$ and its role in the photophysics of molecules.

In the $\ket{T_1;\bm{\nu}_v=\bm{0},\bm{\nu}_p=\bm{0}}$ state, the two unpaired electrons interact through magnetic dipole-dipole interactions, which can be described by the phenomenological model Hamiltonian $\mathbf{H}_\text{ZFS}=\mathbf{S}\cdot\mathbf{D}\cdot\mathbf{S}$, with $\mathbf{S}$ the total spin operator and $\mathbf{D}$ the fine-structure tensor\,\cite{kohler-1999}. For PAH molecules with $D_{2h}$ symmetry, such as pentacene and perylene, $\mathbf{D}$ can be diagonalized in the coordinate system defined by the geometrical symmetry axes $\alpha=x,y,z$ of the molecule, as illustrated with a perylene molecule in Fig.\,\ref{fig:zfs}(a). The three eigenstates associated with the three principal vectors, known as the \textit{zero-field eigenstates}, can be labelled $\ket{T_1^\alpha}$. In $\ket{T_1^{\alpha}}$, the magnetic moment of the electron spin is oriented in the plane perpendicular to the axis $\alpha$. As shown in Fig.\,\ref{fig:zfs}(b), the zero-field splitting is characterized by two parameters $D$ and $E$, which correlate with the delocalization of the spin density and the axial symmetry of the molecule, respectively. Transitions between any pair of the zero-field states are allowed by magnetic-dipole interactions and can be driven by microwave fields\,\cite{kohler-1999}.

In the triplet manifolds, the electron spin couples to nuclear spin wavefunction $\phi_\text{nuc}(\mathbf{I})$ through the hyperfine interaction. This interaction is described by the Hamiltonian $\mathbf{H}_\text{HFI}=\sum_i \mathbf{S}\cdot{\mathbf{A}}^{(i)}\cdot{\mathbf{I}_i}$, where $\mathbf{I}_i$ is the nuclear magnetic moment of the $i^\text{th}$ nucleus, and $\mathbf{A}^{(i)}$ is its hyperfine interaction tensor with the electron spin. In hydrocarbon molecular crystals, nuclear spins mainly originate from protons and $^{13}$C isotopes in the guest molecule and the host matrix. Notably, protons can be replaced by deuterons via isotope exchange reactions, through which the number of spin-1/2 nuclei in the guest and the host can be significantly reduced. Using deuterated matrices makes it possible to resolve the hyperfine splitting due to individual nuclear spins in a single molecule\,\cite{kohler-1995,wrachtrup-1997}. The right panel of Fig.\,\ref{fig:zfs}(b) illustrates the hyperfine splitting with a magnetic field applied along the $z$ axis and considering only a single spin-$1/2$ nucleus in a perylene molecule. Similar to the demonstration with nitrogen-vacancy centers in diamond\,\cite{jelezko-2004}, hyperfine splitting offers the opportunity to selectively access individual nuclear spin states using narrow-band microwave pulses, and facilitate the realization of controlled-rotation gates between electron and nuclear spins. Note that in the case of nitrogen-vacancy centers, having a triplet ground state allows easy access to nuclear spin states. In PAH molecules, hyperfine interaction is only present in the triplet excited states. This feature increases the complexity of accessing the nuclear spins, but at the same time offers an optically controllable switch of the hyperfine interaction.

\begin{figure}[t]
\begin{center}
\includegraphics[width=0.99\columnwidth]{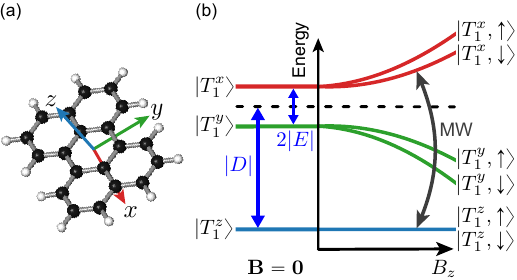}
\caption{(a) Structure of a perylene molecule and its three symmetry axes. (b) Left: Zero-field splitting of the three triplet sublevels $\ket{T_1^{\alpha}}$ in the absence of an external magnetic field ($\mathbf{B}=0$), characterized by parameters $D$ and $E$. Right: Zeeman shift and hyperfine splitting due to a spin-$1/2$ nuclear spin, when an external magnetic field $B_z$ is applied along the $z$ axis. In the regime of a weak magnetic field, the zero-field bases $\ket{T_1^{\alpha}}$ are still approximated eigenstates of the electron spin and $\ket{\uparrow}$, $\ket{\downarrow}$ denote the two states of the nuclear spin. Microwave (MW) driving of the electron spin enables controlled-rotation gates, conditioned on the state of the nuclear spin.}
\label{fig:zfs}
\end{center}
\end{figure}

The triplet state can be populated through optical pumping on the ZPL transition. From the $\ket{S_1;\bm{\nu}_v=\bm{0},\bm{\nu}_p=\bm{0}}$ state, the molecule can occasionally cross over to a vibrationally excited level in the triplet manifold $\ket{T_j;\bm{\nu}_v\neq\bm{0},\bm{\nu}_p\neq\bm{0}}$. This process, known as \textit{intersystem crossing} (ISC), is weakly allowed due to spin-orbital coupling, and typically assisted by intermolecular and intramolecular vibrations\,\cite{penfold-2018}. Following ISC, the molecule undergoes nonradiative relaxation and reaches the lowest triplet state $\ket{T_1;\bm{\nu}_v=\bm{0},\bm{\nu}_p=\bm{0}}$.  The rate of ISC also depends on the final spin sublevel $\ket{T_1^{\alpha}}$ of the triplet manifold\,\cite{turro-2010-book}. From the state $\ket{T_1;\bm{\nu}_v=\bm{0},\bm{\nu}_p=\bm{0}}$, the molecule can decay to the ground state manifold either via phosphorescence, or through ISC to a $\ket{S_0;\bm{\nu}_v\neq\bm{0},\bm{\nu}_p\neq\bm{0}}$ level followed by vibrational relaxation. These two processes compete with each other. The rate of ISC has been shown to follow the energy-gap law\,\cite{wilson-2001}, while its underlying microscopic mechanisms are generally complex\,\cite{beljonne-2001}. Since both processes are spin-forbidden, the lifetime of the $\ket{T_1;\bm{\nu}_v=\bm{0},\bm{\nu}_p=\bm{0}}$ state is long, reaching microseconds to milliseconds\,\cite{Kolchenko-2005, Smit-2022}.

In the triplet state, the molecule does not fluoresce, so a decrease in fluorescence intensity can serve as an indicator of triplet-state occupation. By using a resonant microwave drive to shuffle the population between two of the three sublevels, the mean residence time in the triplet state can be altered, resulting in changes to the steady-state fluorescence intensity. This technique was applied to pentacene molecules in $para$-terphenyl crystal in the 1990s, which led to the first detections of electron spin transitions in a single molecule\,\cite{wrachtrup-1993,kohler-1993}. These efforts were later extended to deuterated pentacene molecules and enabled the detection of the hyperfine splittings due to individual ${}^{13}$C and proton spins\,\cite{kohler-1995,wrachtrup-1997}. Apart from fluorescence detection, spin transitions among the triplet sublevels have also been measured through atomic force microscopy on a single pentacene molecule\,\cite{sellies2023-single}. Very recently, the detection and coherent control of triplet electron spins have been demonstrated on an ensemble of pentacene molecules at room temperature\,\cite{Mena-2024}, potentially setting the ground for ambient quantum sensing using molecular host-guest materials.

\section{Perspectives in Quantum Technology}
\label{sec:qt}

So far, single organic molecules have found applications as single photon sources, quantum optical nonlinear elements, and nanoscale sensors for probing local field properties in materials. They have also been integrated into nanophotonic and hybrid nanostructures, such as nanoantennas, waveguides and optical resonators, laying out promising future paths in photonic quantum technologies\,\cite{toninelli-2021}. The vibrational and spin states of molecular host-guest systems, as discussed in the previous section, hold significant potential for expanding the application of these systems. Furthermore, the chemical diversity of molecules could enhance this potential by providing molecular candidates with tailored quantum properties. In the following, we outline possible strategies to harness the vibrational and spin states for applications that have not been realized on these systems before and discuss the opportunities presented by the vast chemical space.

\subsection{Vibrational quantum memory}\label{sec:Vib_memory}
A quantum memory enables the storage of quantum information and plays a pivotal role in quantum computing and communication~\cite{kimble-2008}. Among the multitude of energy levels in a single molecule, the vibrational degrees of freedom can potentially be harnessed for quantum memories. 

The schematic of a vibrational quantum memory in a single molecule is shown in Fig.~\ref{fig:Fig_QuantumMemory}, where the molecular electronic transition provides the interface to flying qubits and quantum memory is realized in the ground state vibrational manifold. The flying qubits encoded in photon-number~\cite{duan-2001} or time-bin degrees of freedom~\cite{bhaskar-2020} can be stored and retrieved with well-established techniques such as stimulated Raman scattering~\cite{Reim-2010,gurlek-2021} as shown in Fig.~\ref{fig:Fig_QuantumMemory}, stimulated Raman adiabatic passage~\cite{specht-2011} or heralded Stokes Raman scattering~\cite{duan-2001}. 

Although the electronic and vibrational levels of molecules exhibit sizable couplings, the short vibrational lifetimes, which are typically on the picosecond timescale, pose a significant limitation for quantum memory applications. To address this challenge, it has been proposed that GHz-frequency acoustic phonon modes in the host crystal can be exploited as long-lived quantum memory elements~\cite{gurlek-2021}, provided that the molecular environment is carefully engineered. These so-called continuum modes can be made spectrally resolvable by employing nanometer-scale host crystals, which are compatible with current nanofabrication techniques~\cite{hail-2019,pazzagli-2018}. Such small mode volumes further allow for strong electron-phonon coupling strengths that can exceed the decoherence rates of the excited electronic state. Moreover, lifetimes of these acoustic phonon modes can be extended up to the millisecond regime through phononic crystal engineering~\cite{MacCabe-2020}, enabling their use as vibrational quantum memories.

The operation of such a memory is inherently limited to millikelvin temperatures due to thermal phonon occupation but could potentially be extended to higher temperatures by employing bulk acoustic wave resonators~\cite{kharel-2018}. Efficient operation additionally requires near-perfect coupling of single photons to the single molecule. While free-space optics can achieve up to $13\%$ coupling efficiency~\cite{maser-2016}, photonic structures offer values approaching $93\%$~\cite{wang-2019}. Since the quantum memory protocol relies on coherent interactions, both the quantum efficiency of the single molecule and dephasing of the phonon mode limit the achievable fidelity. While single-molecule emitters have typically high quantum efficiency, the fidelity of quantum memory may nonetheless be constrained by the dephasing of phonon modes, which remains insufficiently characterized.

In addition to the acoustic phonon modes, quasi-localized lattice vibrational modes~\cite{Kolchenko2009} and two-level system defects~\cite{Zumbusch1993} in host crystals are promising platforms for quantum memory implementation~\cite{Neeley2008,Li2024}. These excitations exhibit strong coupling to electronic transitions and possess long relaxation times, often reaching up to the millisecond regime, which can be further enhanced at millikelvin temperatures or through phonon engineering techniques~\cite{Chen2024}.

Considering the availability of nanophotonic interfaces, phonon engineering strategies, and the high quantum efficiencies achievable with single molecules, this platform presents a compelling route towards realizing molecular quantum memories.

\begin{figure}[t]
\begin{center}
\includegraphics[width=\columnwidth]{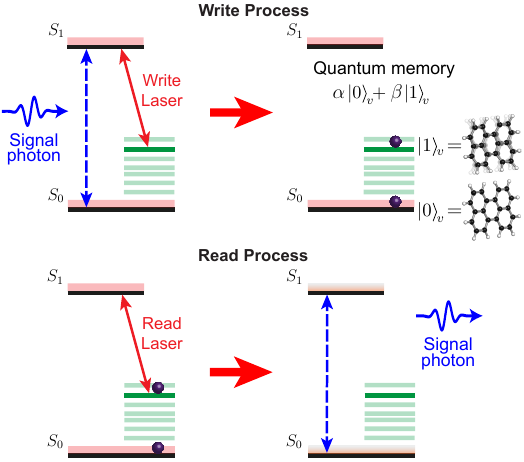}
\caption{Schematics showing photonic quantum memory operation via stimulated Raman scattering process. Flying qubits are interfaced to the ZPL transition ($\ket{S_0;\bm{\nu}_v=\bm{0},\bm{\nu}_p=\bm{0}}-\ket{S_1;\bm{\nu}_v=\bm{0},\bm{\nu}_p=\bm{0}}$) shown with the dashed blue arrows and mapped into (from) the vibrational state with write (read) laser show with the red arrows. The vibrational states of a single molecule in the ground state manifold $\ket{0}_v$ and $\ket{1}_v$ act as a quantum memory. Here, $\ket{0}_v$ and $\ket{1}_v$ are short hand notations for $\ket{S_0;\bm{\nu}_v,\bm{\nu}_p=\bm{0}}$ and $\ket{S_0;[0,\ldots,0,1,0\dots,0],\bm{\nu}_p=\bm{0}}$, respectively. The schema is also valid for phonon states~\cite{gurlek-2021}.}
\label{fig:Fig_QuantumMemory}
\end{center}
\end{figure}

\subsection{Molecular optomechanics}
Optical and mechanical degrees of freedom typically interact weakly in systems beyond the atomic scale. This is due to the large mechanical mass, which makes mechanical systems less responsive to the small momentum of optical fields. In the last decade, highly cooperative interactions between those have been made possible by utilizing engineered optomechanical cavities with low optical and mechanical losses~\cite{barzanjeh-2022}. These efficient interactions enabled precise control and manipulation of mechanical motion with light, and hold promises for quantum sensing, signal transduction, and so on.

Analogous to conventional optomechanical setups, a single-molecule host-guest system offers optomechanical interactions thanks to the inherent couplings between their optically accessible electronic and vibrational transitions. The molecular system provides such couplings across a very wide frequency range spanning kHz to THz. Importantly, the electron-vibration coupling strength ($g_0$) can reach values comparable to the mechanical oscillation frequencies of vibrons~\cite{gurlek-2023-thesis}, i.e., $g_0\sim0.1-0.5~\omega_v$, putting the molecular system into the ultra-strong optomechanical coupling regime at the single-photon level~\cite{aspelmeyer-2014}. This nonlinear quantum optomechanical regime has not been reached in solid-state cavity optomechanical setups so far. 

Together with the extremely large quality factors of the electronic transitions ($\gamma_0\sim$~MHz, where $\gamma_0$ is the spontaneous decay rate of electronic transitions), these ultra-strong coupling strengths enable the system to achieve a high quantum optomechanical cooperativity, $C_\text{opt}$, given by
$C_\text{opt}=4g_0^2/\bar{n}\kappa_v\gamma_0$~\cite{aspelmeyer-2014}. Here, $\bar{n}$ represents the thermal population of the vibrational modes and $\kappa_v$ is the decay rate of the vibrational excited state. Although the vibrons have a short lifetime ($\kappa_v\sim 10~\text{ps}$), the combination of these parameters enables the realization of $C_\text{opt}>10^5$ over a wide range of temperatures ($\bar{n}\ll 1$).
Reaching such a high cooperativity regime in cavity optomechanical systems is challenging due to parasitic optical absorption~\cite{ren-2020}, which is negligible in molecular systems. Hence, single molecules could facilitate quantum optomechanical applications such as coherent frequency conversion between photons~\cite{xomalis-2021}, relevant for quantum information processing~\cite{mirhosseini-2020} and molecular spectroscopy~\cite{Chikkaraddy2023}.

Despite the potential for achieving high quantum optomechanical cooperativity in single-molecule host-guest systems, the inherently short picosecond-scale coherence times of molecular vibrations pose significant challenges for practical applications. These short lifetimes limit the ability to perform coherent manipulations and necessitate the use of high-power laser driving~\cite{Lombardi-2018}, ultrafast excitation and detection schemes or deep-subwavelength plasmonic nanocavities~\cite{xomalis-2021,Chen-2021}. Therefore, extending vibrational lifetimes into the nanosecond regime represents a critical step toward enabling more robust applications.

In traditional cavity optomechanics, similar challenges for acoustic phonons have been addressed using techniques such as impedance mismatch engineering and phononic bandgap structuring, which decouple specific vibrational modes from their environment and thereby prolong coherence. Analogously, in molecular systems, promising routes for achieving efficient optomechanical interactions have been proposed as discussed in Section~\ref{sec:Vib_memory}, even in the presence of modest electron-phonon coupling strengths compared to electron–vibron interactions. However, extending such vibrational control to optical phonons and localized guest-molecule vibrations presents a more formidable challenge due to anharmonic vibrational interactions (see relevant discussions in Section~\ref{label:VibStates}). Potential strategies to tackle this problem include the use of host materials with a sparse vibrational density of states, such as crystalline benzene, the design of molecular systems inherently decoupled from their vibrational surroundings~\cite{corcelli-2002}, or the exploitation of topological degrees of freedom in molecular systems~\cite{Fan2023-topo}. 

Moreover, given their strong electron-vibron coupling and minimal thermal population even at room temperature, engineering long-lived optical phonons and localized
guest-molecule vibrational states could enable interesting quantum optomechanical applications at elevated temperatures~\cite{huang-2024}. Realizing operation above millikelvin temperatures, however, requires not only the suppression of vibrational decay but also the minimization of vibrational dephasing. While negligible dephasing has been demonstrated for certain optical phonon modes in naphthalene molecular crystals up to $60$~K~\cite{Schosser-1984}, similar behavior remains to be explored in single-molecule host-guest systems. This property may offer an advantage over traditional cavity optomechanical setups, which are prone to dephasing even at millikelvin temperatures due to coupling to two level-system defects~\cite{MacCabe-2020}. As a result, the molecular approach offers the potential to explore previously inaccessible regimes in optomechanical and nonequilibrium condensed-matter systems.
\begin{figure}[t]
\begin{center}
\includegraphics[width=\columnwidth]{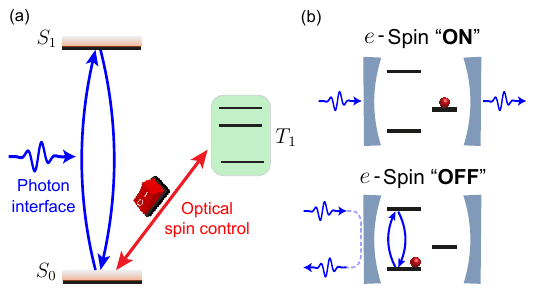}
\caption{(a) Achieving optical manipulation of the singlet-triplet transition (red arrow) will provide a ``switch" to control the electron spin, allowing it to be turned ``on" and ``off" on demand. This capability combined with photon scattering via the $\ket{S_0;\bm{\nu}_v=\bm{0},\bm{\nu}_p=\bm{0}}$ to $\ket{S_1;\bm{\nu}_v=\bm{0},\bm{\nu}_p=\bm{0}}$ ZPL transition (blue arrows) can facilitate optical interface of the electron spin. (b) With an optical cavity resonantly coupled to the ZPL transition, an incoming photon either traverses the cavity when the electron is in the triplet state, or reflects from the front mirror when the electron is in the singlet state.}
\label{fig:Fig_SpinPhoton}
\end{center}
\end{figure}

\subsection{Spin-photon interface}
The spin states are long-lived and well-isolated from their environment, making them attractive candidates as quantum memory units. Additionally, the excellent properties of molecular ZPL emission offer the possibility to coherently interface spins with propagating photons. These two features combined make it possible to realize a molecular spin-photon interface, which can serve as a key component for quantum networks\,\cite{atature-2018}.

To realize an efficient spin-photon interface, it is necessary not only to have optically addressable spin states but also to achieve a coherent coupling of the emitter with a single optical mode. For solid-state emitters, the efficiency of such coherent interaction is constrained by the emitter’s internal quantum efficiency, the branching ratio of ZPL emission, and the optical dephasing rate. Typical PAH molecules used in quantum optical experiments have branching ratio of $\sim$30\% and low intersystem crossing rate ($<10^{-5}$), when a suitable host matrix is chosen. For the near-infrared-emitting dibenzoterrylene molecule, quantum efficiency larger than 50\% has been recently measured\,\cite{Musavinezhad-2024}. While for molecules emitting in the visible range, higher quantum efficiencies are expected, following the energy-gap law. This combination of deficiencies can be mitigated through resonant coupling to an optical cavity. Using a Fabry-Perot cavity with a $Q$-factor of $2.3\times 10^5$ and mode volume of 4.4\,$\lambda^3$, recent experiments have achieved strong cavity-molecule coupling with cooperativity of 45\,\cite{wang-2021,pscherer-2021}. In this regime, emitter-cavity coupling can be combined with spin control to realize a spin-photon interface using deterministic interaction schemes\,\cite{reiserer-2022}.

Figure\,\ref{fig:Fig_SpinPhoton}(a) illustrates a possible scheme of a molecular spin-photon interface. To entangle the electron spin with photons, one needs to prepare the molecule in a superposition of $\ket{S_0;\bm{\nu}_v=\bm{0},\bm{\nu}_p=\bm{0}}$ and a sublevel of $\ket{T_1;\bm{\nu}_v=\bm{0},\bm{\nu}_p=\bm{0}}$. This can in principle be achieved using a narrow linewidth laser to coherently drive the transition between these two states. The spin-photon interaction takes place as follows. As illustrated in Fig.\,\ref{fig:Fig_SpinPhoton}(b), the $\ket{S_0;\bm{\nu}_v=\bm{0},\bm{\nu}_p=\bm{0}}$ to $\ket{S_1;\bm{\nu}_v=\bm{0},\bm{\nu}_p=\bm{0}}$ transition is strongly coupled to an optical microcavity and serves as an optical interface. Depending on the state of the molecule, a photon with its frequency matching the undressed cavity frequency will either reflect off from the cavity, due to the vacuum Rabi splitting\,\cite{pscherer-2021} when the molecule is in $\ket{S_0;\bm{\nu}_v=\bm{0},\bm{\nu}_p=\bm{0}}$, or transmit through the cavity when the molecule is in $\ket{T_1;\bm{\nu}_v=\bm{0},\bm{\nu}_p=\bm{0}}$, considering a symmetric mirror configuration. Thus the internal state of the molecule becomes entangled with the spatial mode of the photon.

A critical step in this scheme is the optical control of singlet-triplet transition, as shown by the red arrow in Fig.\,\ref{fig:Fig_SpinPhoton}(a), which has not yet been demonstrated on a single PAH molecule. Such transitions have been studied in experiments with cold alkaline earth atoms, e.g., strontium, and utilized for optical clocks transitions\,\cite{kyungtae-2023}. In recent years, singlet-triplet transitions have been measured on diatomic molecules in the gas phase\,\cite{truppe-2019, walter-2022}. Compared to the gas-phase experiments, molecules embedded in solids bring along several additional challenges. First, PAH molecules have planar or near-planar structures and do not contain heavy atoms, leading to weak spin-orbit coupling and thus long triplet-state lifetimes. Second, the exact frequencies of these transitions are difficult to predict, and the matrix introduces shifts to the transition frequency compared to that of a molecule in vacuum. Third, the coherent manipulation of singlet-triplet transition needs to surpass deficiencies led by non-radiative decay and pure dephasing resulted from coupling to phonons. Although triplet-state lifetime can be routinely measured, determining other key parameters such as the phosphorescence quantum yield, branching ratio and pure dephasing rate has been rarely reported, apart from an earlier work on perylene in anthracene\,\cite{walla-1998}. A considerable amount of preparatory work in triplet spectroscopy is necessary to unlock the outlined possibilities. In addition, considerations need to be taken in identifying molecular candidates for such investigations. We will discuss possible molecular candidates in Section\,\ref{sec:chemical-space}.

\subsection{Molecular spin register}
A quantum register represents a collection of individual qubits, grouped to serve as a basic component for information processing. The molecular framework offers an attractive route for constructing chemically configurable registers of nuclear spins. Depending on the number of active nuclear spins, two operation regimes can be interesting. First, a small ensemble of $10-20$ protons coupled to a single electron spin, which is naturally provided by a PAH molecule, constitutes a well-defined \textit{central-spin system}. Such a system offers extensive possibilities to harness the collective spin states for quantum memory\,\cite{taylor-2003, denning-2019} and decoherence protection\,\cite{kurucz-2009}. Along this route, highly efficient dynamic polarization of nuclear spins has been achieved on pentacene molecules doped in naphthalene hosts, showing remarkably long relaxation times of over 900 hours\,\cite{eichhorn-2013,Quan-2019}. The know-how in these experiments can be transferred to realize quantum control of collective spin states at the single molecular level. In the second regime, a register of a few addressable spin-1/2 nuclei, realized through e.g., partial deuteration of the molecule, could provide a platform where the state of individual nuclei spins can be detected and manipulated independently. This can be made possible using microwave pulses to perform conditional gates between the electron and nuclear spin, similar to the experiments performed on color centers in diamond\,\cite{jelezko-2004}. It is worth noting that pairs of chemically identical ${}^{13}$C spins in a deuterated pentacene molecule have been spectrally resolved in optically detected magnetic resonance measurements\,\cite{brouwer-1999}. Their significantly narrowed transition linewidth indicates an efficient decoupling from the environment. Such a ``dark'' state of a pair of chemically identical nuclei is highly promising for decoherence-protected encoding schemes\,\cite{reiserer-2016}.

Apart from nuclear spin qubits, using nuclei with larger spin quantum numbers can extend the system to a qudit\footnote{A quantum system with a $d$ number of levels.} register. Nuclear spins with higher multiplicity offer an interesting route for quantum information processing beyond qubit-based approaches. They are predicted to bring along reduced numbers of physical quantum units and gate operations for accomplishing a certain computing task\,\cite{chiesa-2023}. A parallel line of research on molecular qudits is being pursued in the context of molecular nanomagnets\,\cite{chiesa-2023}. On the PAH platform, it is readily feasible to incorporate a defined number of spin-1 nuclei, e.g., {${}^{14}$N, through bay functionalization with imide groups\,\cite{nowak-2019}. Along this line, perylenediimide and terrylenediimide derivatives have already been studied in single-molecule optical spectroscopy\,\cite{mais-1997,lang-2007}.

Overall, the molecular framework provides the capability to integrate electron spin, nuclear spins, and optical interfaces within a single platform. Compared to well-studied platforms such as semiconductor quantum dots, rare-earth ions and diamond color centers, the proposed system based on PAH molecules operates in different regimes in terms of system dimensions and hyperfine coupling strengths. In quantum dot systems, the central spin is surrounded by a large number ($\sim10^5$) of nuclear spins of the atoms constituting the dot with hyperfine interaction strengths stronger than the linewidth of the electron spin resonance~\cite{Warburton-2013}. In this regime, which may be termed the strongly-coupled \textit{dense-spin limit}, the constituent nuclear spins are indistinguishable to the central spin. The control over individual nuclear spin states is out of reach but collective states of the ensemble can be probed and addressed~\cite{Urbaszek-2013}. Recent advances in rare-earth ion systems, operated in a comparable parameter regime, have demonstrated that a subset of lattice spins can be isolated from the dense ensemble through the frozen-core effect, enabling quantum state preparation within an isolated subset~\cite{Ruskuc-2022}. In contrast, color centers in diamond, such as the nitrogen-vacancy center, provide an implementation of central-spin systems in the \textit{dilute-spin limit}. The color center can have a few strongly coupled nuclear spins from the nitrogen nucleus forming the center and ${}^{13}\text{C}$ atoms on nearby lattice sites, with hyperfine coupling strengths ranging from 300\,kHz to 130\,MHz~\cite{awschalom-2018}. In addition, weakly coupled nuclear spins, i.e., spins with hyperfine couplings narrower than the electron spin resonance linewidth can also be detected through dynamic decoupling sequences. This enables selective addressing of individual nuclear spins, due to their smaller number and different hyperfine coupling strengths. While it offers advantages in coherence time and controllability, limitations arise from the fact that the majority of addressable nuclear spins are weakly coupled and randomly dispersed within the diamond lattice.

In a PAH molecule, the hyperfine interaction strength of proton and ${}^{13}\text{C}$ spins significantly exceeds the electron spin transition linewidth of $\sim 120\,\text{kHz}$~\cite{kohler-1995}, putting them in the strongly-coupled regime. The molecular framework further allows the spin composition to be varied through chemical modifications, such as perdeuteration. In addition to the dopant molecules, the solid-state environments for PAH molecules, which are normally crystals of hydrocarbon molecules of a smaller size, can also be configured. For example, the transition of a dense-spin to a dilute-spin environment can be achieved through deuteration of the matrix molecules~\cite{kohler-1999}, allowing tuning of the coupling strength between the spin register and its environment. In addition to the tailorability by chemical modification, a particularly appealing aspect of this system is the metastable triplet state. Once optical driving of such a transition is realized, it enables a deterministic control to switch ``on" and ``off" the electron spin [see the red arrow in Fig.~\ref{fig:Fig_SpinPhoton}(a)]. This feature is particularly compelling, given that the electron spin acts as a significant source of decoherence for the nuclear spins~\cite{awschalom-2018}. By enabling a deterministic on-off switch of the hyperfine interaction, this system can address the trade-off between gate time and susceptibility to decoherence, further enhancing the utility of the molecular platform.

\begin{figure*}[t]
\centering
\includegraphics[width=17.8cm]{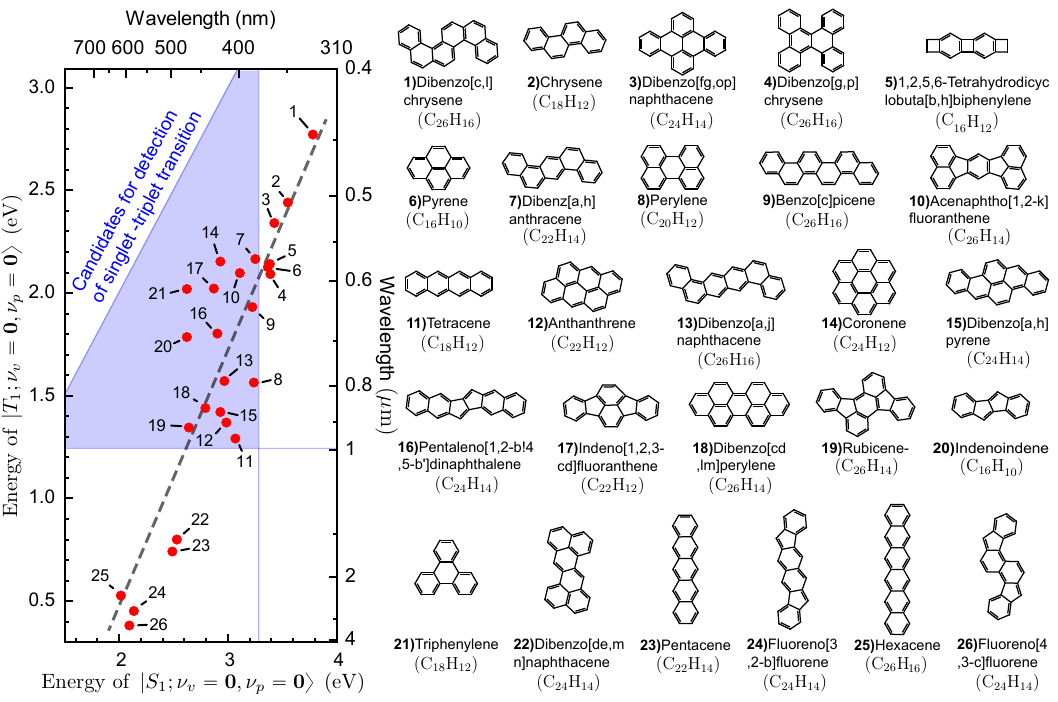}
\caption{The computed vertical excitation energies of $\ket{S_1;\bm{\nu}_v=\bm{0},\bm{\nu}_p=\bm{0}}$ and $\ket{T_1;\bm{\nu}_v=\bm{0},\bm{\nu}_p=\bm{0}}$ of all centrosymmetric PAHs  containing up to $26$ carbon atoms found in the NIST Chemistry WebBook~\cite{linstrom-2001}. The ground state geometry optimizations are performed at the DFT level with B3LYP functional~\cite{Lee1988,Becke1993,Stephens1994}, def2-TZVP basis set~\cite{def2TZVP} with its auxiliary basis set~\cite{def2TZVPC} and the D3 dispersion correction~\cite{D3_1,D3_2}. The vertical excitation energies are calculated using DLPNO-STEOM-CCSD method with def2-TZVP basis set~\cite{berraud-2019,sanyam-2023}. The molecular candidates suitable for the optical detection of $\ket{S_0;\bm{\nu}_v=\bm{0},\bm{\nu}_p=\bm{0}}-\ket{T_1;\bm{\nu}_v=\bm{0},\bm{\nu}_p=\bm{0}}$ transition are highlighted by the blue-shaded area. The grey dashed line is a guide to an eye for the linear scaling between vertical excitation energies. The simulations are carried out via ORCA program, v.$5$.$0$.~\cite{neese-2022}.}
\label{fig:chemical-variety}
\end{figure*}

\subsection{Uncovering the Chemical Space}
\label{sec:chemical-space}
Having introduced the potential of vibrational and spin states within the molecular host-guest system, we now extend the discussion to another research line that remains largely unexplored, namely the molecular candidates within the broader chemical space. Previous achievements in molecular quantum optics have relied on a handful of molecule-host combinations that are identified based on intuition and processes of trial and error. The chemical space of molecules involves more than a billion species~\cite{sanchez-2018}, offering enormous potential to provide molecules with a wide range of desirable properties.

We provide a glimpse into the chemical space of molecules in Fig.\,\ref{fig:chemical-variety}, where $26$ centrosymmetric PAH molecules containing up to $26$ carbon atoms are identified from the NIST database~\cite{linstrom-2001}. Their singlet-singlet ($\ket{S_0;\bm{\nu}_v=\bm{0},\bm{\nu}_p=\bm{0}}-\ket{S_1;\bm{\nu}_v=\bm{0},\bm{\nu}_p=\bm{0}}$) and singlet-triplet ($\ket{S_0;\bm{\nu}_v=\bm{0},\bm{\nu}_p=\bm{0}}-\ket{T_1;\bm{\nu}_v=\bm{0},\bm{\nu}_p=\bm{0}}$) vertical transition energies are calculated using \textit{ab initio} methods at the level of coupled-cluster theory with single and double excitations (CCSD)~\cite{berraud-2019} and plotted versus each other. Note that these energies do not include solvation effects or zero-point energy correction stemming from the vibrational degrees of freedom. The results show an approximately linear scaling behavior between the singlet-singlet and singlet-triplet transition energies\,(see the dashed black line in Fig.\,\ref{fig:chemical-variety}), spanning a large range in the visible and near-infrared spectrum. This molecular space has not been studied at the single-molecule level except for perylene and pentacene. We discuss the potential of such a diverse chemical space with two possible applications.

(i) The large span of $\ket{S_0;\bm{\nu}_v=\bm{0},\bm{\nu}_p=\bm{0}}-\ket{S_1;\bm{\nu}_v=\bm{0},\bm{\nu}_p=\bm{0}}$ transition energies of these molecules offers possibilities to engineer single photon sources with on-demand emission frequencies across the visible spectrum. Although the identified molecules do not cover the entire range, the wavelength gaps among them could be bridged using several strategies. First, chemical modifications, such as the adaptation of imide or anhydride groups, and bay-position functionalization can change the ZPL emission wavelength over an extended range\,\cite{mais-1997,lang-2007}. Second, selecting different host matrices or different insertion sites in a host can affect the emission wavelength by up to $\sim$20\,nm\,\cite{adhikari-2023}. Third, within a given host matrix, the center of inhomogeneous broadening can be tuned with the thickness of the host crystal, providing an accessible frequency range of several nanometers\,\cite{gmeiner-2016}. Fourth, the inhomogeneous broadening of ZPL frequencies can span over $\sim$1\,nm, allowing spectral selection of molecules within this frequency range. Finally, fine adjusting of the ZPL emission frequency can be achieved through the Stark effect, with a tunning range of $\sim$500\,GHz\,\cite{Duquennoy-2024}. Together, these strategies constitute a toolbox for engineering molecular emitters with emission wavelengths tailorable on demand.

(ii) The molecules explored in Fig.~\ref{fig:chemical-variety} offer a valuable chemical space for identifying molecular candidates suitable for realizing spin-photon interfaces, where the optical addressing of the singlet-triplet transition is essential but remains an ongoing challenge~\cite{adhikari-2023}. Molecules suitable for addressing singlet-triplet transitions should meet two conditions regarding their transition energy: First, molecules should have a high $\ket{T_1;\bm{\nu}_v=\bm{0},\bm{\nu}_p=\bm{0}}$ energy, as dictated by the energy-gap law, meaning that a lower energy of $\ket{T_1;\bm{\nu}_v=\bm{0},\bm{\nu}_p=\bm{0}}$ is associated with a larger non-radiative ISC rate~\cite{wilson-2001}. Second, the energy of $\ket{S_1;\bm{\nu}_v=\bm{0},\bm{\nu}_p=\bm{0}}$ should not exceed visible frequencies significantly, as this would complicate the identification of suitable host materials and necessitate the use of widely tunable laser sources in the ultraviolet, which is technically demanding. Since the energy of $\ket{T_1;\bm{\nu}_v=\bm{0},\bm{\nu}_p=\bm{0}}$ is typically lower than that of $\ket{S_1;\bm{\nu}_v=\bm{0},\bm{\nu}_p=\bm{0}}$, it is crucial to identify molecules that can balance these energy levels to facilitate efficient spin-photon interactions. These two criteria constrain the selection of suitable guest molecules to the subspace illustrated by the blue-shaded area in Fig.\,\ref{fig:chemical-variety}.

These two examples are just a glimpse of the immense possibilities offered by chemical space. In the future, significantly larger regions of this space may be explored through machine learning approaches in both theoretical frameworks~\cite{koczor-2021} and experimental workflows~\cite{abolhasani-2023}, enabling the identification of new host-guest combinations with specific properties such as ZPL frequencies at telecom wavelengths, long vibrational lifetimes, strong electron–vibration coupling, extended spin coherence times or high quantum yield emissions from molecular transitions. Organometallic complexes, for example, have already demonstrated the potential to achieve ZPL frequencies at telecom wavelengths~\cite{laorenza-2022}, showcasing the capabilities that could be realized as more chemical space is explored. In addition, inverse molecular design
techniques~\cite{sanchez-2018} can be employed to design new molecules with tailored photonic, vibronic, and spin characteristics, harnessing the full potential of chemical synthesis. These approaches can be integrated with advanced synthetic methodologies, such as click chemistry~\cite{Kolb2001} and iterative molecular assembly~\cite{Lehmann-2018}, to enable the realization of custom-designed molecular systems. Ultimately, leveraging generative models and autonomous laboratories could push the boundaries of this chemical space exploration~\cite{abolhasani-2023,Anstine-2023}. These strategies hold the promise to transcend traditional design paradigms and significantly accelerate the development of molecules for quantum technologies.

\section{Open challenges}
\label{sec:challenges}
Several challenges need to be addressed both in theory and in experiments to unlock the potential of molecular host-guest systems discussed above. Here, we iterate on three major challenges that we identify from state-of-the-art understandings.

First, advancing solid-state molecular quantum technology requires a deeper understanding of molecule–host interactions from first principles. Static interactions, such as the insertion of a single molecule into the host matrix, which influence ZPL frequencies and vibrational properties, can often be addressed through full or semi-classical simulations~\cite{mcrae-1957,bordat-1998,nicolet-2007,bertoni-2022}. However, current studies remain far from being quantitative due to the inherent inhomogeneity of local molecular environments~\cite{zirkelbach-2021}. Meanwhile, dynamical interactions including spectral diffusion, pure dephasing, non-radiative relaxations, and fluorescence blinking pose additional challenges, as they critically affect the optical properties of single molecules in the solid state by introducing decoherence~\cite{Basche-2008-book}. These interactions not only hinder the utilization of all molecular degrees of freedom but also constrain the choice of suitable host materials. A comprehensive understanding and characterization of such interactions are therefore key to the development of controllable and scalable molecular quantum technologies. However, modelling these interactions is particularly challenging due to the involvement of multiple degrees of freedom and their wide energy scales. Addressing this multi-scale problem requires a multidisciplinary approach that integrates tools from condensed matter theory and physical chemistry, including first-principles calculations and model systems. These theoretical efforts must also be supported by dedicated experiments to reveal the microscopic dynamics governing these systems. In the near term, these efforts may face significant challenges due to computational complexity, which could ultimately necessitate quantum computing resources. However, emerging machine-learning technologies provide a promising avenue to tackle these complexities. For instance, deep learning strategies can be employed to extract insights from the vast datasets generated by joint experimental and theoretical efforts~\cite{Gebhart2023}, enabling the prediction of molecular properties and their interactions with the environment. Beyond enhancing our understanding of microscopic dynamics, this line of research could also enable quantum control in molecular quantum technologies~\cite{Koch2022}, opening new possibilities for engineering molecular systems with specific applications. 

A second major challenge lies in the material domain. Host materials for single molecules, which so far have been almost exclusively limited to organic materials apart from a recent demonstration with hexagonal boron nitride (hBN)\,\cite{smit-2023}, often lack physical and chemical stability. Photostable single molecules are typically found in crystalline materials with long-range order. This is normally produced through sublimation in vacuum or inert gas atmosphere. Single molecules in polymeric hosts such as polyethylene\,\cite{rattenbacher-2023} and polymethyl methacrylate\,\cite{walser-2009-spectral} have also been reported, however, with sub-optimal photostability. Recently developed fabrication methods including liquid phase reprecipitation\,\cite{pazzagli-2018,schofield-2022} and electrohydrodynamic nanoprinting\,\cite{hail-2019} could produce crystals with sizes in the micrometer to sub-micrometer scale, and enable Fourier-limited emission~\cite{Musavinezhad-2024}. However, these hosts are fragile due to their low melting temperature, making them prone to sublimation. Additionally, their solubility in common organic solvents makes them hardly compatible with standard lithographic processing steps. To address this challenge, efforts in experiment and theory should be joined to identify an essential and adequate set of criteria to benchmark host-guest combinations. The search for physically and chemically stable, large bandgap host materials within the vast chemical space could benefit from integrating computational molecular screening techniques. High-throughput methods and machine learning-based approaches can efficiently predict the stability, electronic properties, and compatibility of potential host materials, enabling identification of promising candidates~\cite{Ferrenti-2020}. The recent discovery of significant line narrowing of single molecules on the surface of hBN\,\cite{smit-2023} opens a promising prospect to extend the chemical space of host materials beyond organics.

In addition to identifying new host materials, the deterministic integration of single molecules into nanoscale photonic and electronic devices is essential for achieving scalability in these technologies. A third challenge lies in the positioning of single molecules with the desired accuracy. Various approaches have been explored so far. For example, sparsely distributed nanocrystals could be identified through optical microscopy, and photonic structures can be fabricated directly around the selected molecule through direct laser writing\,\cite{colautti-2020b}. Alternatively, sublimated flakes of the host material can be picked up using a fiber tip and transferred onto a desired location on a substrate using micro-manipulation\,\cite{ren-2022}. Electrohydrodynamic nanoprinting has demonstrated the capability to place molecules on a substrate with lateral precision of approximately 100\,nm\,\cite{hail-2019}. Despite these advances, deterministic positioning of individual molecules with atomic-scale precision and assembly of multiple molecules into ordered arrays are still outstanding challenges. It is thus crucial to develop new methodologies for deterministically positioning molecules within the material. Notably, such deterministic positioning and ordering have been demonstrated in surface science experiments, for example, through tip manipulation in a scanning-tunneling microscope (STM)\,\cite{luo-2019,zhang-2017}. However, the conditions of surface science experiments differ considerably with respect to those of quantum optics, from the required material substrate to vacuum conditions and to optical imaging techniques. Technical challenges need to be addressed to bridge the experimental settings of these two fields. Two-dimensional (2D) materials \cite{smit-2023,wang2024-twodimensional} present a promising solution, potentially serving as substrates suitable for both STM manipulations and quantum optical measurements. For example, nanoscopic positioning and alignment of molecules on a 2D-material substrate can be performed on an STM. The substrate can be transferred for stacking of van-der-Waals hetero-structures and fabrication of quantum devices.

\section{Discussions and conclusion}
\label{sec:discussion}
So far, we have introduced the various degrees of freedom in the molecular guest-host system, elaborated on their interactions, and discussed their potential applications in advancing quantum technologies. Specifically, we focused on the vibrational and spin degrees of freedom, and examined the feasibility of constructing vibrational quantum memory, spin-photon interfaces, molecular spin registers, and investigation of molecular optomechanics.

In addition to these paths, it is worth noting that the narrow ZPL of PAH molecules makes them highly sensitive probes of their local physical environment, such as electric fields and strain. Frequency tuning of the ZPL transition via the DC Stark effect has been routinely performed on single molecules. The demonstrated high electric field sensitivity of DBT molecules in 2,3‐dibromonaphthalene matrix theoretically enables the detection of single electrons in the vicinity of the molecule\,\cite{moradi-2019}. This capability opens a range of possibilities for interfacing with quantum electronic devices, such as superconducting circuits\,\cite{das-2017}, electron-based qubit devices in semiconductors\,\cite{langrock-2023, burkard-2023} and 2D materials\,\cite{hecker-2023}. Furthermore, it will also be worth investigating molecules with a permanent dipole moment for providing higher charge sensitivity\,\cite{faez-2015}. In addition, the large strain susceptibilities of these molecules could also be utilized to interface with other quantum systems. For example, molecular spin degrees of freedom~\cite{laorenza-2022} could potentially benefit from such phononic couplings, to achieve couplings through a phononic bus ~\cite{kuzyk-2018}. Such phononic platforms could be realized with polymer host matrices feasible for nanostructuring~\cite{colautti-2020b}, printing nanocrystals~\cite{hail-2019}, or placing single molecules on conventional phononic structures~\cite{gurlek-2021}. In a similar vein, pseudolocal vibrational modes could offer new ways to entangle molecules located at intermediate distances, bridging the electronic and phononic length scales and paving the way for the creation of collective many-body states.

Looking forward, single molecules as hybrid systems offer a range of potential quantum applications. They come with unique flexibility in host-guest combinations, allowing chemical tuning over the quantum mechanical properties of both the guest and the host. Despite challenges in theoretical understandings and fabrication techniques, we expect molecular host-guest systems to play a significant role in future solid-state quantum technologies.

\section*{Acknowledgement}
We thank Michael Ruggenthaler for his valuable comments on the manuscript, and Leonardo dos Anjos Cunha for his help analyzing CCSD simulation results. D.W. acknowledges financial support from Germany’s Excellence Strategy - Cluster of Excellence Matter and Light for Quantum Computing (ML4Q) EXC 2004/1-390534769 and the European Union (ERC, MSpin, 101077866).

\section*{Data Availability}
The data that support the findings of this article are openly available~\cite{gurlek_2025_15323133}.

\end{document}